\documentclass
[aps,prl,twocolumn,showpacs,floatfix,superscriptaddress]{revtex4-1}%
\usepackage{graphicx}
\usepackage{amsmath}
\usepackage{physics}
\usepackage{amssymb}
\usepackage{colordvi}
\usepackage{verbatim}
\usepackage{xcolor}
\usepackage{mathrsfs}
\usepackage{epsfig}
\usepackage{lipsum}
\usepackage{amsfonts}%
\providecommand{\U}[1]{\protect\rule{.1in}{.1in}}
\providecommand{\U}[1]{\protect\rule{.1in}{.1in}}
\setcitestyle{numbers,square}

\begin{document}

\title{Unidirectional Pumping of Phonons by Magnetization Dynamics}
\author{Xiang Zhang}
\affiliation{Kavli Institute of NanoScience, Delft University of Technology,
	2628 CJ Delft, The Netherlands}
\author{Gerrit E. W. Bauer}
\affiliation{Institute for Materials Research \& WPI-AIMR \& CSRN, Tohoku
	University, Sendai 980-8577, Japan} \affiliation{Kavli Institute of
	NanoScience, Delft University of Technology, 2628 CJ Delft, The Netherlands}
\author{Tao Yu}
\email{tao.yu@mpsd.mpg.de}
\affiliation{Max Planck Institute for the
	Structure and	Dynamics of	Matter,	22761 Hamburg, Germany}
\affiliation{Kavli Institute of NanoScience, Delft University of Technology,
	2628 CJ Delft, The Netherlands}
\date{\today }

\begin{abstract}
We propose a method to control surface phonon transport by weak magnetic
fields based on the pumping of surface acoustic waves (SAWs) by
magnetostriction. We predict that the magnetization dynamics of a nanowire on
top of a dielectric films injects SAWs with opposite angular momenta into
opposite directions. Two parallel nanowires form a phononic cavity that at
magnetic resonances pump a unidirectional SAW current into half of the substrate.

\end{abstract}
\maketitle

\textit{Introduction.}---Surface acoustic waves (SAWs) on the surface of
high-quality piezoelectric crystals are frequently employed for traditional
signal processing \cite{Ash2014,Kino1987}, but are also excellent mediators
for coherent information exchange between distant quantum systems such as
superconducting qubits and/or nitrogen-vacancy centers
\cite{Gustafsson2014,Manenti2017,Satzinger2018,Golter2016}. Piezoelectrically
excited coherent SAWs drive the ferromagnetic resonance (FMR) by
magnetostriction
\cite{Delsing2019,Weiler2011,Weiler2012,SAW_pumping,Yahagi2014,Sasaki2017,Otani2020}%
, excite spin waves parametrically \cite{nonlinear_PRB}\textit{,} and generate
electron spin currents by the rotation-spin coupling \cite{Matsuo2013,SRC_PRL}%
. Conventional insulators often have good acoustic quality but only small
piezoelectric effects, rendering the direct excitation, manipulation and
detection of the coherent SAWs challenging. The phonon pumping
\cite{Streib2018}, i.e., the excitation of bulk sound waves in a high-quality
acoustic insulator by the dynamics of a proximity magnetic layer via the
magnetoelastic coupling \cite{Holanda2018,Berk2019}, may be useful here. Bulk
phonons in the insulator gadolinium gallium garnet (GGG) can couple two
yttrium iron garnet (YIG) magnetic layers over millimeters
\cite{An2020,utrecht}.

\begin{figure}[th]
\centering
\includegraphics[width=6.8cm]{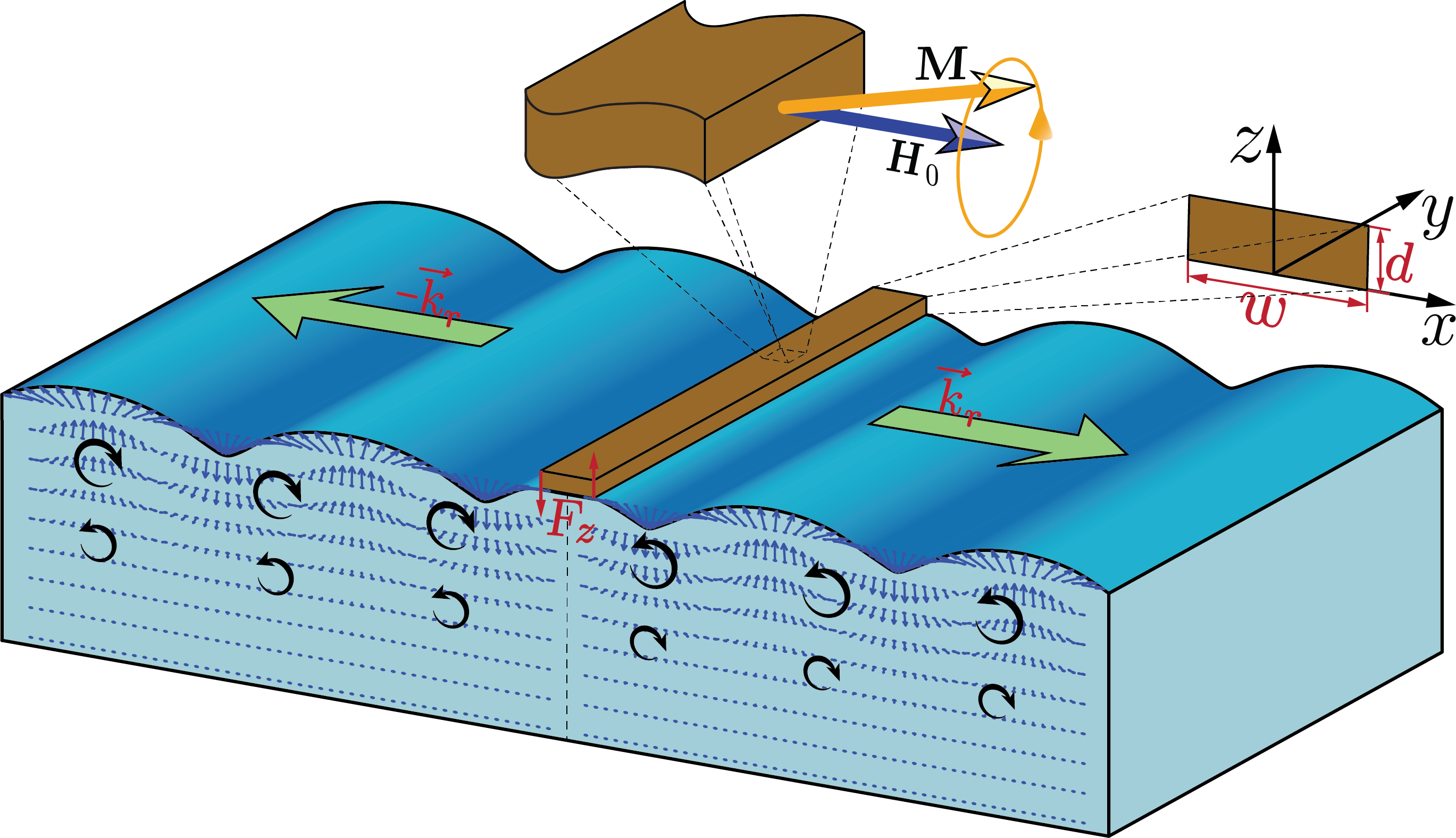}\caption{Surface-phonon pumping by one
magnetic nanowire (brown) on top of the acoustic insulator (blue). A static
magnetic field $\mathbf{H}_{0}$ applied in the $\Vec{x}$-direction saturates
the magnetization. The pumped Rayleigh SAWs by the nanowire FMR propagate and
rotate in opposite directions at the two sides of the nanowire as indicated by
the green and black arrows, respectively.}%
\label{fig:setup}%
\end{figure}

Here we address the coherent excitation and manipulation of Rayleigh-SAWs by
magnetization dynamics, which is possible in a lateral planar configuration
with ferromagnetic nanowires on top of a high-quality nonmagnetic insulator,
as illustrated in Fig.~\ref{fig:setup}. Similar configurations on magnetic
substrates led to the electrical detection of diffuse magnon transport
\cite{Cornelissen2015,Das} and discovery of nonreciprocal magnon propagation
\cite{Haiming}, i.e. the generation of a unidirectional spin current in half
space \cite{Yu1,Tao2019}. Magnetic stray fields of the magnetization dynamics
also generate chiral electron \cite{chiral_electron} and waveguide photon
\cite{chiral_photon} transport. The unidirectional excitation of SAWs is
important for acoustic device applications \cite{acoustic_book}, which
conventionally is achieved by metal electrodes on a piezoelectric crystal such
that reflected SAWs constructively interfer with the source. This is a pure
geometrical effect that is efficient at sub-GHz frequencies and sample
dimensions that match the SAW wavelength \cite{acoustic_book,SPUDT}.

We focus on the\textit{ unidirectional} \cite{Note} excitation of SAWs via
magnetic nanostructures on top of a dielectric substrate that are brought into
FMR by external microwaves. We predict effects that are very different from
the reported nonreciprocity, i.e. a sound velocity that depends on direction
\cite{Camley,Sasaki2017}, which is enhanced in magnetic multilayers on top of
a piezoelectric substrate \cite{slavin1,slavin2,Otani_anisotropic}. The
magnetic order of, e.g., a wire on top of a dielectric, does not couple
nonreciprocally to the surface phonons in the configuration in
Fig.~\ref{fig:setup}, but excites both left- and right-propagating phonons,
even though the angular momentum current has a direction because of the
momentum-rotation coupling of Rayleigh SAWs. However, we predict robust
unidirectional excitation of SAW phonons in a phononic cavity formed by two
parallel wires. The SAWs actuated by the first wire interact with the second
one (which does not see the microwaves) and excite its magnetization, which in
turn emits phonons. The phonons from both sources interfere destructively on
half of the surface and the net phonon pumping becomes unidirectional.
Constructive interference between the two nanowires induces standing SAWs as
in a Fabry-P\'{e}rot cavity. Conventional unidirectional electric transducer
\cite{acoustic_book,SPUDT} operate by a pure geometrical interference effect
that works only for a fixed sub-GHz frequency. Our magnetic unidirectional
transducer operates by a dynamical phase shift and provides new
functionalities, such as robust high frequency tunability and switchability.

\textit{Model.---}We consider a rectangular magnetic nanowire (YIG) on top of
the surface of a thick dielectric (GGG) that spans the $x,y$ plane. It extends
along the $y$-direction with $z\in\left[  0,d\right]  $ and $x\in
x_{i}+\left[  -w/2,w/2\right]  $, as shown in Fig.~\ref{fig:setup}. For an
analytical treatment, $d$ is assumed to be much smaller than the skin depth of
the SAWs, such that the displacement field in the wire is nearly uniform in
the $z$-dependence. The lattice and elastic parameters at the YIG$|$GGG
interface match well \cite{elasticity,utrecht,Streib2018,An2020} and are
assumed equal. A uniform and sufficiently large static magnetic field
$\mathbf{H}_{0}$ along $\Vec{x}$ saturates the equilibrium magnetization
$\mathbf{M}_{0}=M_{s}\Vec{x}$ normal to the wire. We can modulate the
magnon-phonon coupling simply by rotating $\mathbf{H}_{0}$.

The system Hamiltonian consists of the elastic energy $\hat{H}_{\mathrm{e}}$,
the magnetoelastic coupling $\hat{H}_{\mathrm{c}},$ and the magnetic energy of
the Kittel mode \cite{Landau1984},
\begin{equation}
\hat{H}_{\mathrm{m}}=\int d\boldsymbol{r}\,\left(  -M_{x}H_{0}+\frac{1}%
{2}N_{xx}M_{x}^{2}+\frac{1}{2}N_{zz}M_{z}^{2}\right)  ,
\label{eqn:Hamiltonian_M}%
\end{equation}
where $\mathbf{M}=(M_{x},M_{y},M_{z})^{T}$ is the magnetization vector and the
demagnetization constants are taken as $N_{xx}\simeq d/(d+w)$ and
$N_{zz}\simeq w/(d+w)$ \cite{Yu1}. Although the predicted effects are
classical, we use a quantum description for convenience and future
applications in quantum phononics
\cite{Schuetz2015,Gustafsson2014,Manenti2017,Satzinger2018,Golter2016}; we can
always recover the classical picture by replacing operators by amplitudes. The
transverse magnetization is quantized by the Kittel-magnon operator
$\hat{\beta}(t)$ with normalized wavefunction $m_{y,z}$
\cite{Walker,surface_roughness}, see Supplemental Material \cite{supplemental}%
:
\begin{equation}
\hat{\mathbf{M}}_{y,z}=-\sqrt{2\gamma\hbar M_{s}}\left(  m_{y,z}\hat{\beta
}(t)+m_{y,z}^{\ast}\hat{\beta}^{\dagger}(t)\right)  ,
\label{eqn:magnon_operator}%
\end{equation}
leading to $\hat{H}_{\mathrm{m}}=\hbar\omega_{\mathrm{F}}\hat{\beta}^{\dagger
}\hat{\beta}$ with frequency
\[
\omega_{\mathrm{F}}=\mu_{0}\gamma\sqrt{(H_{0}-N_{xx}M_{s})(H_{0}-N_{xx}%
M_{s}+N_{zz}M_{s})}.
\]
Here, $-\gamma$ and $\mu_{0}$ are the gyromagnetic ratio and vacuum permeability.

In our configuration, only the Rayleigh SAWs couple efficiently with the
magnet which by their surface nature and long decay length are well suited to
exchange information with spatially remote magnets (see Supplemental Material
\cite{supplemental}). Sufficiently thin and narrow wires do not affect the
substrate strongly, so we may treat them perturbatively. The surface
eigenmodes of an isotropic elastic half space read \cite{Viktorov1967}
\begin{equation}%
\begin{split}
\psi_{x}  &  =ik\varphi_{k}\left(  e^{qz}-\frac{2qs}{k^{2}+s^{2}}%
e^{sz}\right)  e^{ikx},\\
\psi_{z}  &  =q\varphi_{k}\left(  e^{qz}-\frac{2k^{2}}{k^{2}+s^{2}}%
e^{sz}\right)  e^{ikx},
\end{split}
\label{eqn:SAW_profile}%
\end{equation}
where $q=\sqrt{k^{2}-k_{l}^{2}}$ and $s=\sqrt{k^{2}-k_{t}^{2}}$ with
$k_{l}=\omega_{k}\sqrt{\rho/(\lambda+2\mu)}$ and $k_{t}=\omega_{k}\sqrt
{\rho/\mu}$ are the wave vectors for longitudinal and transverse bulk waves,
respectively. Here, $\rho$ is the material density, $\mu$ and $\lambda$ are
the elastic Lam\'{e} constants, and $\varphi_{k}$ is a normalization constant
\cite{supplemental}. $\omega_{k}=\left\vert k\right\vert \eta\sqrt{\mu/\rho
}=c_{r}|k|$ represents the eigenfrequency of Rayleigh SAWs with velocity
$c_{r}$ and $\eta<1$ is the root of the SAW characteristic equation
\cite{Viktorov1967} that depends only on the Lam\'{e} constants. The relative
phase of the displacement field ${\mathrm{Arg}}(\psi_{z}/\psi_{x})|_{z=0}=\pm
i$ is opposite for right- and left-propagating waves, which reflects the
rotation-momentum locking \cite{Viktorov1967}.

The quantized displacement field $(\hat{u}_{x},\,\hat{u}_{z})$ can be expanded
into the eigenmodes $\boldsymbol{\psi}(k)$ and phonon operators $\hat{b}%
_{k}(t)$ \cite{Schuetz2015}
\begin{equation}
\hat{\mathbf{u}}(x,z,t)=\sum_{k}\left[  \boldsymbol{\psi}(x,z,k)\hat{b}%
_{k}(t)+\boldsymbol{\psi}^{\ast}(x,z,k)\hat{b}_{k}^{\dagger}(t)\right]  .
\label{eqn:phonon_operator}%
\end{equation}
We normalize the mode amplitudes $\boldsymbol{\psi}$ to recover the elastic
Hamiltonian for Rayleigh SAWs \cite{supplemental} such that
\begin{equation}
\hat{H}_{\mathrm{e}}=\rho\int d\boldsymbol{r}\,\dot{\hat{\boldsymbol{u}}}%
^{2}(x,z,t)=\sum_{k}\hbar\omega_{k}\hat{b}_{k}^{\dagger}\hat{b}_{k}.
\label{eqn:Hamiltonian_E}%
\end{equation}

In YIG films the magnetocrystalline anisotropy, which is important in CoFeB
\cite{Otani2020,shear_strain}, is relatively weak
\cite{Streib2018,elasticity,utrecht} and the (linearized) magnon-phonon
coupling energy is dominated by $\hat{H}_{c}^{m}=(B_{\perp}/{M_{s}}%
)\int\left(  \hat{M}_{y}\partial_{x}u_{y}+\hat{M}_{z}\partial_{z}u_{x}+\hat
{M}_{z}\partial_{x}u_{z}\right)  d\boldsymbol{r}$ with magnetoelastic coupling
constant $B_{\perp}$ \cite{Landau1984,Hillebrands2014}. The magnetic wire and
non-magnetic substrate are coupled by the dynamics of the surface strain. We
require the interaction between a given SAW and the Kittel mode. By the
translational symmetry along the nanowire $y$-direction, the displacement
field excited by the Kittel magnon does not depend on $y$, and SAWs
propagating along $x$ do not contribute to $u_{y}$. The magnetoelastic energy
contributed by the magnetic wire with length $L$ then becomes
\begin{align}
\hat{H}_{c}^{m}  &  =\frac{B_{\perp}L}{M_{s}}\int_{0}^{d}\hat{M}_{z}\left(
u_{z}|_{x={w} /{2}+x_{i}}-u_{z}|_{x=-{w}/{2}+x_{i}}\right)  dz\nonumber\\
&  +\frac{B_{\perp}L}{M_{s}}\int_{-{w}/{2}+x_{i}}^{{w} /{2}+x_{i}}\hat{M}%
_{z}\left(  u_{x}|_{z=d}-u_{x}|_{z=0}\right)  dx. \label{eqn:interaction}%
\end{align}
We limit attention to the realistic situation in which the wire thickness $d$
is much smaller than the decay length of the SAWs into the bulk. The strain in
the magnet then mirrors that of the SAW at $z=0$ of the dielectric and the
second term in Eq.~(\ref{eqn:interaction}) vanishes \cite{Otani2020},
\begin{equation}
\hat{H}_{c}^{m}\rightarrow\frac{B_{\perp}Ld}{M_{s}}\hat{M}_{z}\left(
u_{z}|_{x={w}/{2}+x_{i}}-u_{z}|_{x=-{w}/{2}+x_{i}}\right)
\label{interaction2}%
\end{equation}
that corresponds to an oscillating surface force $\boldsymbol{F}|_{x=\pm
{w}/{2}+x_{i}}=\mp{B_{\perp}}Ld\hat{M}_{z}/{M_{s}}\Vec{z}$ in the
$z$-direction that excites SAWs traveling outwards in both directions
\cite{supplemental}. Substituting Eqs.~(\ref{eqn:magnon_operator}) and
(\ref{eqn:phonon_operator}) into Eq.~(\ref{interaction2}), we arrive at the
interaction Hamiltonian
\begin{equation}
\hat{H}_{c}=\hbar\sum_{k}g_{k}\hat{\beta}^{\dagger}\hat{b}_{k}+\mathrm{H.c.},
\end{equation}
in which the coupling constant ($qd\ll1$, $sd\ll1$)
\begin{equation}
g_{k}\simeq-B_{\perp}\sqrt{\frac{\gamma}{M_{s}\rho c_{r}}}\sqrt{\frac{d}{w}%
}\sin\left(  \frac{kw}{2}\right)  \xi_{\mathrm{M}}\xi_{\mathrm{P}}e^{ikx_{i}},
\label{eqn:coupling_constant}%
\end{equation}
with factors $\xi_{\mathrm{M}}$ and $\xi_{\mathrm{P}}$ governed by the
magnetic and acoustic material parameters \cite{supplemental}. The form factor
oscillates and decreases algebraically as a function of nanowire width and
phonon wavelength and vanishes when $d,w\rightarrow0$. The coupling is
\emph{reciprocal} since $\left\vert g_{k}\right\vert =|g_{-k}|$.

\textit{SAW pumping.}---We now calculate the phonon pumping by a single
magnetic nanowire transducer centered at $x_{0}$ and excited by microwave
photons represented by the (annihilation) operator $\hat{p}_{\mathrm{in}}$.
The Hamiltonian $\hat{H}=\hat{H}_{\mathrm{m}}+\hat{H}_{\mathrm{e}}+\hat{H}%
_{c}$ leads to the Heisenberg equation of motion
\cite{Gardiner1985,Clerk2010}
\begin{align}
{d\hat{\beta}}/{dt}  &  =-i\omega_{\mathrm{F}}\hat{\beta}-i\sum_{k}%
|g_{k}|e^{ikx_{0}}\hat{b}_{k}-\left(  {\kappa_{m}}+{\kappa_{\omega}}\right)
\hat{\beta}/2\nonumber\\
&  -\sqrt{\kappa_{\omega}}\hat{p}_{\mathrm{in}},\nonumber\\
{d\hat{b}_{k}}/{dt}  &  =-i\omega_{k}\hat{b}_{k}-i|g_{k}|e^{-ikx_{0}}%
\hat{\beta}-{\delta_{k}}\hat{b}_{k}/{2}, \label{eqn:EOM}%
\end{align}
where $\kappa_{m}$ and $\delta_{k}$ are the intrinsic damping rates for the
nanowire magnon and surface phonon, while $\kappa_{\omega}$ is the radiative
damping induced by the microwave field. In the frequency domain,
$\hat{\mathcal{O}}(\omega)=\int dt\hat{\mathcal{O}}(t)e^{i\omega t}$,
\begin{align}
\hat{\beta}(\omega)  &  =\frac{-i\sqrt{\kappa_{\omega}}\hat{p}_{\mathrm{in}%
}(\omega)}{\omega-\omega_{\mathrm{F}}+i(\kappa_{m}+\kappa_{\omega})/2-\sum
_{k}|g_{k}|^{2}G_{k}(\omega)},\nonumber\\
\hat{b}_{k}(\omega)  &  =G_{k}(\omega)|g_{k}|e^{-ikx_{0}}\,\hat{\beta}%
(\omega), \label{eqn:solve_EOM}%
\end{align}
where $G_{k}(\omega)=1/(\omega-\omega_{k}+i\delta_{k}/2)$ is the phonon Green
function. The additional magnetic damping by the phonon pumping at the FMR
\cite{Streib2018,Streib2019} is given by the imaginary part of the magnon
self-energy
\begin{equation}
\sigma_{k}(\omega)=-\mathrm{\operatorname{Im}}\left(  \sum_{k}|g_{k}|^{2}%
G_{k}(\omega)\right)  =\frac{|g_{k_{r}}|^{2}}{c_{r}}, \label{eqn:damping}%
\end{equation}
where we use the on-shell approximation \cite{Mahan,Vignale} with
$\omega\rightarrow\omega_{\mathrm{F}}$ and $k_{r}=\omega_{\mathrm{F}}/c_{r}$.
The real part of the self-energy causes a small frequency shift that is
absorbed into $\omega_{\mathrm{F}}$ in the following.

The displacement field given by Eq.~(\ref{eqn:phonon_operator}) is a
superposition of coherent phonons $\langle\hat{b}_{k}\rangle$ that are excited
by the microwave input $\left\langle \hat{p}_{\mathrm{in}}(\omega
)\right\rangle $. At resonance $\omega\rightarrow\omega_{\mathrm{F}}$, the
contour of the $k$ integral must be closed in the\ upper (lower) half of the
complex plane for $x>x_{0}$ ($x<x_{0}$), selecting the poles $k_{r}%
+i\epsilon\ $($-k_{r}-i\epsilon$) in the denominator, where $\epsilon$ is the
inverse of the phonon propagation length. The low ultrasonic attenuation in
GGG at room temperature corresponds to characteristic SAW decay lengths of up
to 6~mm \cite{Dutoit1974}. We can therefore safely disregard the phonon
damping ($\epsilon\rightarrow0_{+}$), which leads to displacement fields
\begin{equation}
\mathbf{u}(z,t)=-\frac{2}{c_{r}}\mathrm{\operatorname{Re}}\left\{
\begin{array}
[c]{c}%
i\boldsymbol{\psi}(k_{r},z)g_{k_{r}}^{\ast}\langle\hat{\beta}(t)\rangle
,~~~~~~x>x_{0}\\
i\boldsymbol{\psi}(-k_{r},z)g_{-k_{r}}^{\ast}\langle\hat{\beta}(t)\rangle
,~~x<x_{0}%
\end{array}
\right.  .
\end{equation}
On the right (left) side of the nanowire $x>x_{0}$ ($x<x_{0}$), the right- and
left-propagating waves with opposite rotations, whose directions depend on
$z$, are pumped as illustrated in Fig.~\ref{fig:setup}. A classical treatment
leads to the same result \cite{supplemental}.

These phonons carry a constant mechanical angular momentum density
$\boldsymbol{l}_{\mathrm{DC}}\left(  x,z\right)  =\rho\left\langle
\mathbf{u}\times\mathbf{\dot{u}}\right\rangle _{t}$, where the subscript $t$
indicates time average, which is often referred to as \textit{phonon spin}
\cite{Long2018,Holanda2018,Shi2019}:
\begin{align}
\boldsymbol{l}_{\mathrm{DC}}(z)  &  =(4\rho\omega_{\mathrm{F}}/c_{r}%
^{2})|\langle\hat{\beta}\rangle|^{2}|g_{k_{r}}|^{2}\Vec{y}\nonumber\\
&  \times\mathrm{\operatorname{Im}}\left\{
\begin{array}
[c]{c}%
\psi_{x}(k_{r},z)\psi_{z}^{\ast}(k_{r},z),~~~~~~~x>x_{0}\\
\psi_{x}(-k_{r},z)\psi_{z}^{\ast}(-k_{r},z),~~x<x_{0}%
\end{array}
\right.  . \label{excited_u}%
\end{align}
$\boldsymbol{l}_{\mathrm{DC}}$ is proportional to the excited magnon
population, parallel to the wire, and opposite on both sides of the nanowire
since $\psi_{x}(-k)\psi_{z}^{\ast}(-k)=-\psi_{x}(k)\psi_{z}^{\ast}(k)$. Into
the substrate ($z$-direction), the SAW eigenmodes\ have a node at which
$\boldsymbol{l}_{\mathrm{DC}}$ changes sign \cite{supplemental} as sketched in
Fig. \ref{fig:setup}. The phonon pumping does not remove angular momentum from
the ferromagnet, since only the $x$-component of the magnetic precession is
damped. The force on the interface is a superposition of opposite angular
momenta $2\mathbf{z}=(\mathbf{z}+i\mathbf{x})+(\mathbf{z}-i\mathbf{x})$ that
by the spin-momentum locking couple to phonons moving in opposite direction.

\begin{figure}[th]
\hspace{-0.1cm}\includegraphics[width=4.15cm]{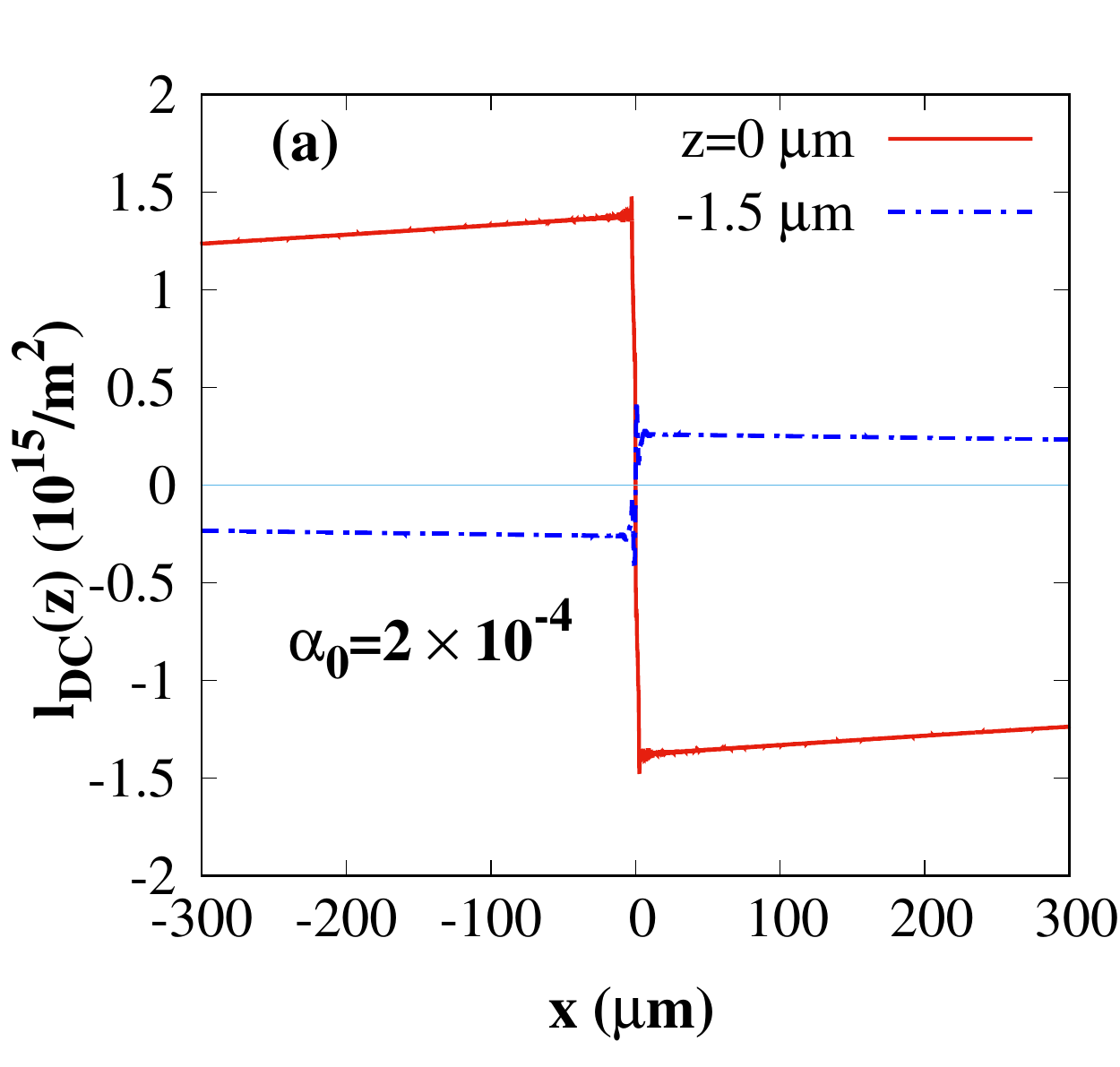}
\hspace{-0.05cm}\includegraphics[width=4.15cm]{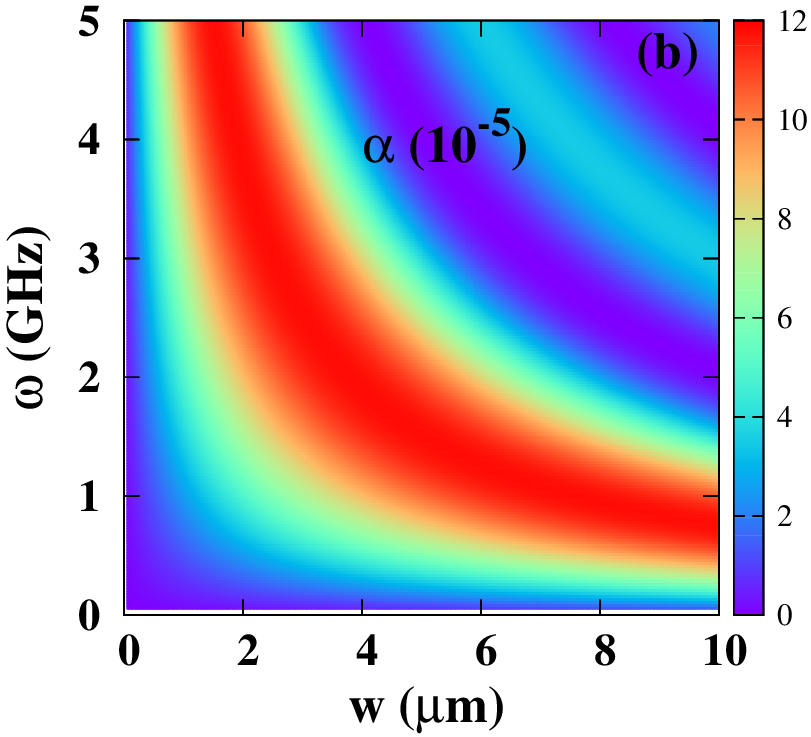}\caption{Pumped
phonon spin density [(a)] and additional magnon damping coefficient $\alpha$
[(b)] for a YIG wire on a GGG substrate as a function of FMR frequencies and
wire widths.}%
\label{fig:damping}%
\end{figure}

The efficiency of phonon spin pumping depends on the nanowire and substrate.
For GGG at room temperature, $\rho=7080\,\mathrm{{kg/m^{3}}}$, $c_{l}%
=6545\,\mathrm{m/s}$ and $c_{t}=3531\,\mathrm{m/s}$ \cite{Schreier2013},
leading to \cite{Viktorov1967} $\eta=0.927$, $c_{r}=\eta c_{t}=3271.8$%
~$\mathrm{m/s}$, and $\xi_{\mathrm{P}}=0.537$. For YIG \cite{YIG_nanowire},
$\gamma=1.82\times10^{11}$~$\mathrm{{s^{-1}T^{-1}}}$, $\mu_{0}M_{s}=0.177$~T
\cite{Serga2010}, $B_{\perp}=6.96\times10^{5}$~$\mathrm{{J/m^{3}}}$
\cite{Streib2018} and $\xi_{\mathrm{M}}\approx1$ when $H_{0}$ is comparable to
$M_{s}$. We plot the pumped phonon spin density at different $z$ in
Fig.~\ref{fig:damping}(a) with $\omega_{\mathrm{F}}=3$~GHz, $d=200~\mathrm{nm}%
$ and $w=2.5~\mathrm{\mu}$m. We use a small precession cone angle $10^{-3}$
degrees and phonon diffusion length $\sim6$~mm. The spin density is opposite
at the two sides of the nanowire and changes sign at larger $z$.
Figure~\ref{fig:damping}(b) is a plot of the additional magnon damping
coefficient $\alpha=\sigma_{k}/\omega_{\mathrm{F}}$ in the dependence of FMR
frequency $\omega_{\mathrm{F}}$ and nanowire width $w$. We observe geometric
resonances $\sim1/w$ with $\alpha\lesssim1\times10^{-4},$ which is of the
order of the intrinsic Gilbert damping of YIG\ single crystals $\alpha_{0}%
\sim4\times10^{-5}$ \cite{Yu_CP} and films $8\times10^{-5}$ \cite{YIG_damping}%
. In the thin YIG film, the additional damping $\alpha\sim d$.

\textit{Unidirectional phonon pumping.}--- The single wire emits spin-momentum
locked SAWs into two directions. We propose a truly unidirectional phonon
source in the form of two parallel and identical nanowires located at
$\mathbf{r}_{1}=R_{1}\vec{x}$ and $\mathbf{r}_{2}=R_{2}\vec{x}$, of which only
the left one is addressed by a local microwave stripline
\cite{transducer_Dirk}. The excited phonons below propagate to and are
absorbed by the second nanowire. Its dynamics re-emits phonons that
subsequently interfere with the original ones \cite{An2020}. Denoting the
magnon operators in the left and right nanowires as $\hat{\beta}_{\mathrm{L}}$
and $\hat{\beta}_{\mathrm{R}}$ \cite{supplemental},
\begin{align}
\hat{\beta}_{\mathrm{R}}(\omega)  &  =\frac{\sum_{k}|g_{k}|^{2}G_{k}\left(
\omega\right)  e^{ik(R_{2}-R_{1})}}{\omega-\omega_{\mathrm{F}}+i\kappa
_{m}/2-\sum_{k}|g_{k}|^{2}G_{k}\left(  \omega\right)  }\hat{\beta}%
_{\mathrm{L}}(\omega),\nonumber\\
\hat{b}_{k}(\omega)  &  =|g_{k}|G_{k}(\omega)\left(  e^{-ikR_{1}}\hat{\beta
}_{\mathrm{L}}(\omega)+e^{-ikR_{2}}\hat{\beta}_{\mathrm{R}}(\omega)\right)  .
\label{eqn:excitations_two}%
\end{align}
At the FMR $\omega\rightarrow\omega_{\mathrm{F}}$,
\begin{equation}
\hat{\beta}_{\mathrm{R}}(\omega_{\mathrm{F}})=\chi(k_{r})e^{i\pi+ik_{r}%
(R_{2}-R_{1})}\hat{\beta}_{\mathrm{L}}(\omega_{\mathrm{F}}),
\label{phase_relation}%
\end{equation}
where $\chi(k_{r})=\sigma(k_{r})/(\kappa_{m}/2+\sigma(k_{r}))$ modulates the
magnetization amplitude in the second wire and $k_{r}(R_{2}-R_{1})$ is the
phase delay by the phonon transmission. The phase shift $\pi$ reflects the
dynamical phase relation between magnons and phonons that is the key for the
unidirectionality. This relation can be observed inductively in microwave
transmission spectra \cite{supplemental}.

By substituting Eq.~(\ref{phase_relation}) into (\ref{eqn:excitations_two}) at
the FMR:
\begin{align}
\hat{b}_{-k_{r}}  &  =|g_{k_{r}}|G_{k_{r}}e^{ik_{r}R_{1}}\hat{\beta
}_{\mathrm{L}}(\omega_{\mathrm{F}})\left(  1-\chi(k_{r})e^{2ik_{r}(R_{2}
-R_{1})}\right)  ,\nonumber\\
\hat{b}_{k_{r}}  &  =|g_{k_{r}}|G_{k_{r}}e^{-ik_{r}R_{1}}\hat{\beta
}_{\mathrm{L}}(\omega_{\mathrm{F}})\left(  1-\chi(k_{r})\right)  .
\end{align}
In the strong magnon-phonon coupling limit $\sigma(k_{r})\gg\kappa_{m}/2$,
$\chi(k_{r})\rightarrow1$, thus the right-going phonon $k_{r}>0$ is not
excited by the double-wire configuration. Finite $\langle\hat{b}_{-k_{r}%
}\rangle$ but vanishing $\langle\hat{b}_{k_{r}}\rangle$ implies a
unidirectional phonon current. Such unidirectionality vanishes when the second
wire is weakly coupled to the SAW, i.e. $\sigma(k_{r})\ll\kappa_{m}/2$, i.e.
phonons transmit without interacting with the magnet.

\begin{figure}[th]
\hspace{-0.18cm}\includegraphics[width=4.25cm]{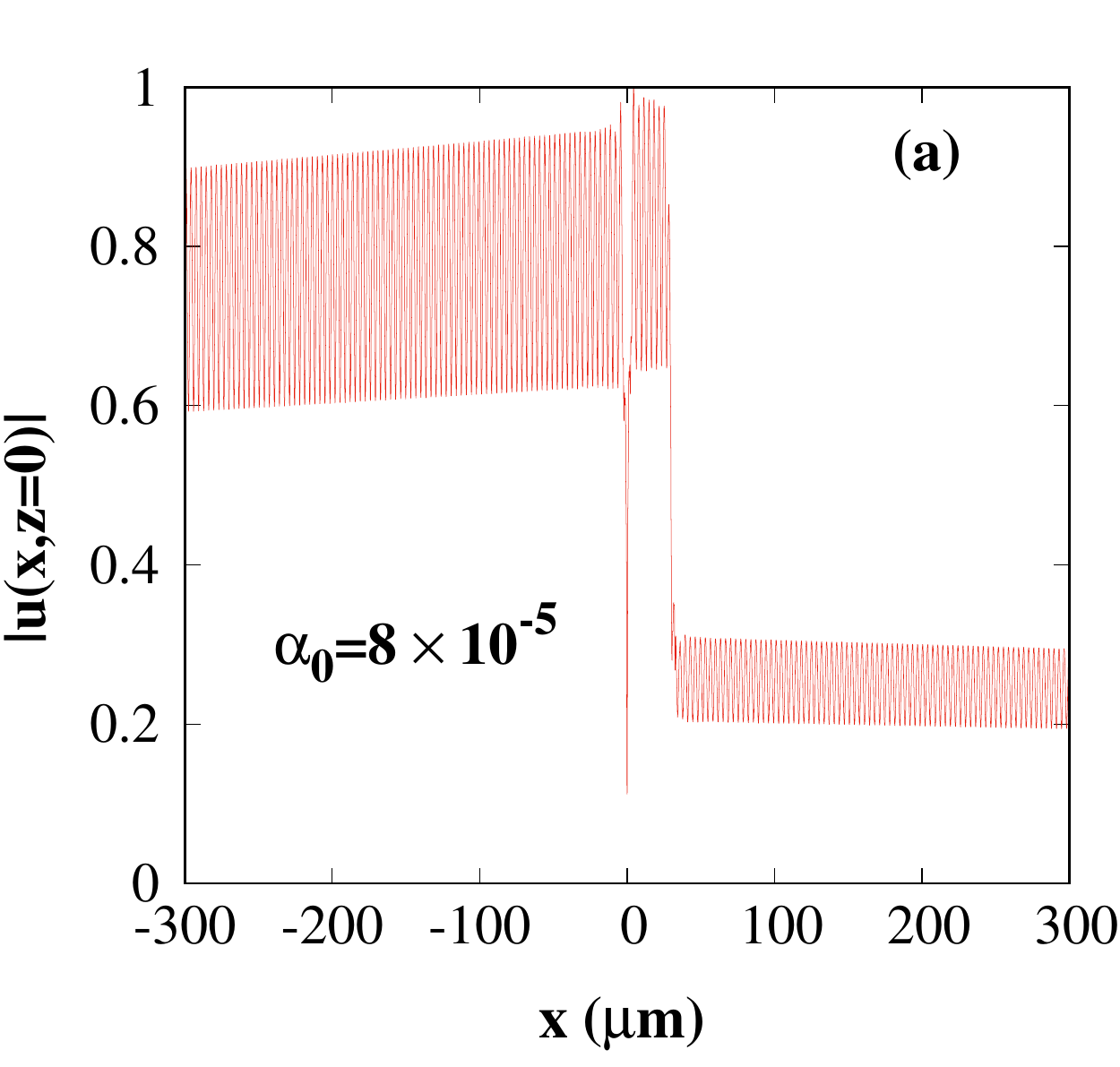}
\hspace{-0.18cm}
\includegraphics[width=4.25cm]{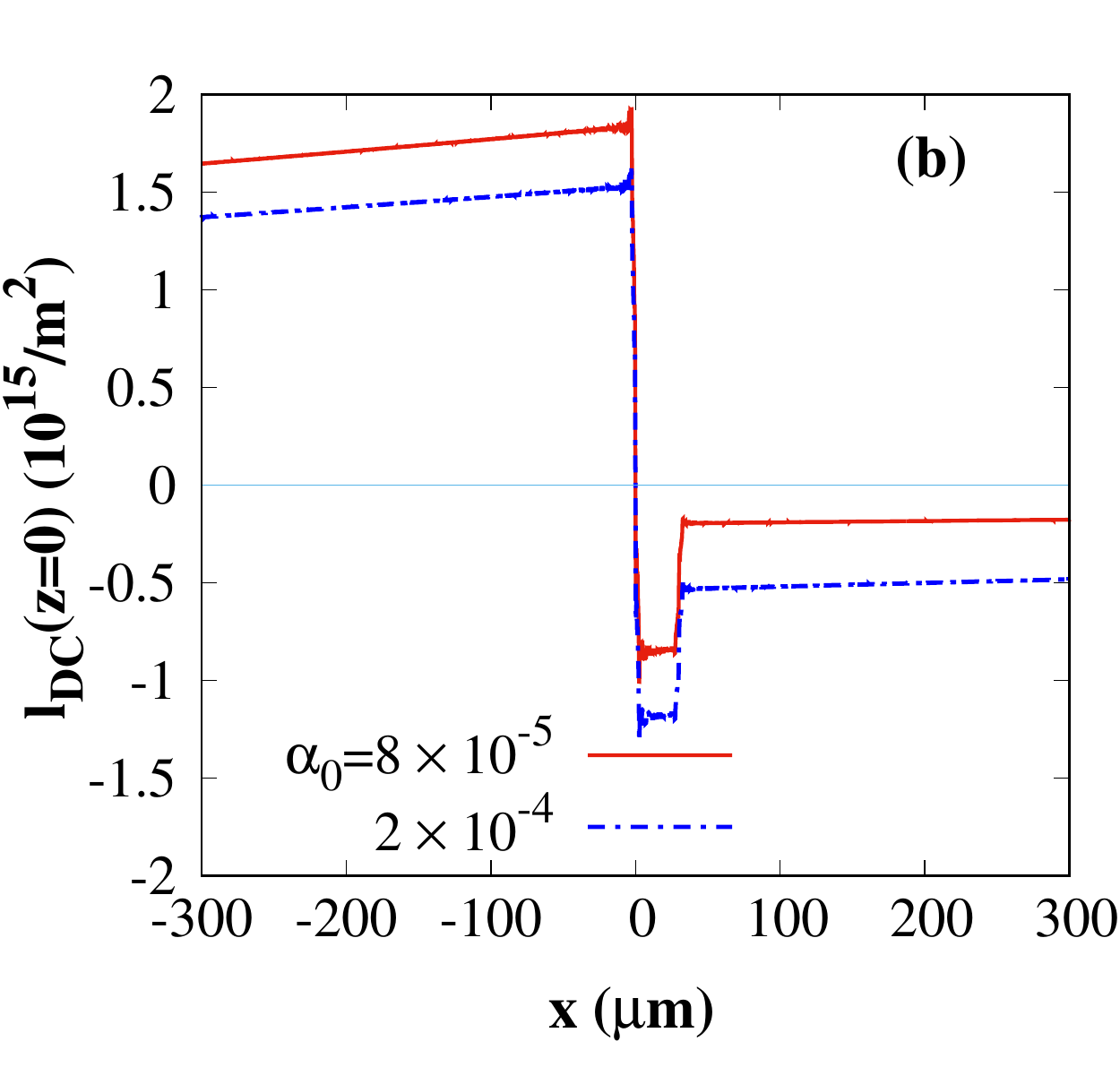}\caption{Snapshot of the
displacement field at the GGG surface $\left\vert \mathbf{u}\right\vert $
[(a)] and phonon spin density [(b)], pumped by a YIG wire at the origin under
FMR and modulated by a second YIG\ wire at $30\,\mathrm{\mu m}$.}%
\label{phonon_spin_density}%
\end{figure}

\begin{widetext}
	By Eqs.~(\ref{eqn:phonon_operator}) and (\ref{eqn:excitations_two}), the
	displacement fields of frequency $\omega_{\mathrm{F}}$ read
\begin{align}
\mathbf{u}(x,t)=\frac{2|g_{k_{r}}|}{c_{r}}\mathrm{Im}\left\{
\begin{array}
[c]{c}%
\boldsymbol{\psi}({-k_{r}})e^{ik_{r}R_{1}}\left\langle \hat{\beta}%
_{L}(t)\right\rangle \left(  1-\chi(k_{r})e^{2ik_{r}(R_{2}-R_{1})}\right)  \\
e^{ik_{r}(R_{2}-R_{1})}\left\langle \hat{\beta}_{L}(t)\right\rangle \left(
\boldsymbol{\psi}({k_{r}})e^{-ik_{r}R_{2}}-\chi(k_{r})\boldsymbol{\psi}%
({-k_{r}})e^{ik_{r}R_{2}}\right)  \\
\boldsymbol{\psi}({k_{r}})e^{-ik_{r}R_{1}}\left\langle \hat{\beta}%
_{L}(t)\right\rangle \left(  1-\chi(k_{r})\right)
\end{array}
~~\text{for}~~%
\begin{array}
[c]{c}%
x<R_{1}\\
R_{1}<x<R_{2}\\
x>R_{2}%
\end{array}
\right.  .
\end{align}
\end{widetext}When $\chi(k_{r})\rightarrow1$, the displacement field vanishes
in the region $x>R_{2}$, but is a traveling wave for $x<R_{1}$. Between the
two nanowires with $R_{1}<x<R_{2}$, the SAWs form standing waves with
$u_{z}\sim\sin{k_{r}(x-R_{2})}$ and $u_{x}\sim\cos{k_{r}(x-R_{2})}$. The
pumping is unidirectional apart from special cases: with frequency or distance
$k_{r}(R_{2}-R_{1})=n\pi$ with $n\in\mathbb{Z}_{0}$, the SAWs on the left hand
side vanishes as well and the phonon is fully trapped between the two wires to
form a cavity. The phonon emission is not perfectly unidirectional when
$\chi(k_{r})<1,$ however. Figure~\ref{phonon_spin_density}(a) is a plot of the
magnitude of the displacement field at the GGG surface $\left\vert
\mathbf{u}(x,z=0)\right\vert =\sqrt{u_{x}^{2}+u_{z}^{2}}|_{z=0}$ with an
intrinsic Gilbert damping $\alpha_{0}=8\times10^{-5}$ \cite{YIG_damping}. We choose
the YIG wires of $d=200$~nm and $w=2.5~\mathrm{\mu}$m to center at $R_{1}=0$
and $R_{2}=30~\mathrm{\mu}$m, and the Kittel frequency $\omega_{\mathrm{F}}%
=3$~GHz such that the additional damping coefficient is $1.2\times10^{-4}$. In
Fig.~\ref{phonon_spin_density}(b) we plot the phonon (DC) spin density at the
GGG surface for a precession cone angle $10^{-3}$ degrees in the left wire.
The asymmetry of the pumped phonon spin at the two sides of YIG cavity is not
perfect, but clearly survives a larger damping. In YIG$\vert$GGG systems,
phonon pumping should be measurable even in less than perfect samples. Devices
with more than two wires or made from magnetic materials with larger
magnetoelasticity may achieve full unidirectionality.

The pumped phonon (spin) can propagate coherently over millimeters on the
substrate surface, which is very promising for classical and quantum transport
of spin information. It can be measured by Brillouin light scattering
\cite{Holanda2018}, the spin-rotation coupling by fabricating a conductor on
top of the acoustic medium \cite{Matsuo2013,SRC_PRL}, and other techniques
\cite{Shi2019}.
The generation of unidirectionality by interference does not require a
nonreciprocal coupling mechanism
\cite{Yu1,Tao2019,chiral_electron,chiral_photon} but only an out-of-phase
relation of the two fields at resonance. The phenomenon should be universal
for many field propagation phenomena, such as exchange coupled magnetic
nanowires and films \cite{Yu1} and reciprocally coupled magnons and waveguide
photons \cite{chiral_photon}.

\textit{Discussion.}---In conclusion, we developed a theory for pumping SAWs
and proposed a phonon cavity device that realizes unidirectional phonon
current in a reciprocal system. When exciting a single magnetic nanowire by
microwaves, we predict a passive wire induces the unidirectional phonon
current and formation of standing waves in the region between two magnetic
nanowires. This mechanism should also lead to a unidirectional\textbf{ }spin
Seebeck effect \cite{Tao2019} generated by a temperature gradient between the
magnetic and acoustic insulators \cite{spin_caloritronics}. Unidirectionality
emerges here from a dynamical phase shift, rather than the purely geometrical
interference employed by electrical unidirectional SAW generators
\cite{acoustic_book,SPUDT}. In the strong coupling regime \cite{An2020} two
magnetic wires on a simple dielectric form a fully unidirectional SAW phonon
source/transducer that opens intriguing perspectives for magnonics and
spintronics, but also plasmonics \cite{Nori_Science,Nori_review}, nano-optics
\cite{nano_optics}, and quantum communication.

\vskip0.25cm \begin{acknowledgments}
	This work is financially supported by the Nederlandse Organisatie voor Wetenschappelijk Onderzoek (NWO) as well as JSPS KAKENHI Grant No. 19H006450.
\end{acknowledgments}


\begin{thebibliography}{67}
\bibitem{Note} In the literature different terms are used to
emphasize different aspects of the same thing. \textquotedblleft
Unidirectional\textquotedblright\ refers to the current direction,
\textquotedblleft chirality\textquotedblright\ is a fundamental symmetry
property that governs this direction, while \textquotedblleft
nonreciprocity\textquotedblright\ is an asymmetry of mutual interactions in
coherent systems by broken symmetries, of which unidirectionality is an
extreme consequence.
\end{thebibliography}
\end{document}